\documentclass[11pt,manuscript]{aastex}
\input psfig.sty

\newcommand {\ea} {{\it et~al.}}
\newcommand {\be} {\begin{equation}}
\newcommand {\ee} {\end{equation}}

\slugcomment{Revised version May 15, 2002}
\shorttitle{MeV blazars}
\shortauthors{Sikora \ea}

\begin{document}

\title{On the Nature of MeV-blazars}

\author{M.~Sikora, M.~B{\l }a\.zejowski, R.~Moderski}
\affil{Nicolaus Copernicus Astronomical Center, Bartycka 18, 00-716
Warsaw, Poland}
\email{sikora@camk.edu.pl}

\and

\author{G.~M.~Madejski }
\affil{Stanford Linear Accelerator Center, 2575 Sand Hill Road, Menlo Park, CA 94025, USA}
\begin{abstract}

Broad-band spectra of the FSRQ (flat-spectrum-radio quasars) detected 
in the high energy gamma-ray band imply that there may be two types 
of such objects:  those with steep gamma-ray spectra, hereafter called 
MeV-blazars, and those with flat gamma-ray spectra, GeV-blazars.  
We demonstrate that this difference can be explained in the context 
of the ERC (external-radiation-Compton) model using the same electron 
injection function.  A satisfactory unification is reachable,
provided that: (a) spectra of GeV-blazars are produced by internal shocks 
formed at the distances where cooling of relativistic electrons in a jet is 
dominated by Comptonization of broad emission lines, whereas spectra of 
MeV-blazars are produced at the distances where cooling of relativistic 
electrons is dominated  by Comptonization of near-IR radiation from hot dust; 
(b) electrons are accelerated via a two step process and their injection
function takes the form of a double power-law, with the break corresponding to 
the threshold energy for the diffusive shock acceleration.  Direct predictions
of our model are that, on average, variability time scales of the MeV-blazars 
should be longer than variability time scales of the GeV-blazars, 
and that both types of the blazar phenomenon can appear in the same object. 

\end{abstract}

\section{INTRODUCTION}

The data obtained from the Compton Gamma-ray Observatory (CGRO) mission 
suggest that blazars -- the subclass of AGNs which includes FSRQ and BL Lac 
objects -- are strong $\gamma$--ray emitters (von Montigny et 
al. 1995;  Mukherjee et al. 1997).  In these objects $\gamma$--ray 
radiation forms a distinctive spectral component, clearly separated from 
another, lower energy component, presumably produced by the synchrotron 
process.  As it was demonstrated by Fossati et al. (1998), blazar spectra 
form a sequence which can be parametrized by their total luminosity.  
In this sequence, FSRQs are the most luminous objects.  Both their low and 
high energy spectral components appear to be the least extended 
to their respective 
high energies, and their $\gamma$--ray luminosity during flares strongly 
dominate over the luminosity of the synchrotron component.  
The least luminous blazars are represented by the X--ray selected BL Lac 
objects, often called HBLs (High-energy peaked BL Lacs).  Their low energy 
spectral component extends up to hard X--rays, and their $\gamma$--ray 
spectrum reaches TeV energies.  $\gamma$--ray flux detected in the 
TeV-emitting BL Lac objects usually does not dominate over synchrotron flux. 

As predicted already in 1978 by Blandford and Rees and later supported
by many independent observations and theoretical analyses, blazar 
radiation is most likely produced by nonthermal plasma in 
relativistic jets and Doppler boosted in our direction.
Due to this, the Doppler enhancement of the jet renders the thermal 
components of the nucleus such as the UV radiation from an accretion 
disc, the X--ray radiation from the disc corona, and the infrared 
radiation from the hot dust -- all presumably emitted isotropically -- 
significantly diluted.  But, at least in FSRQs, the optical-UV broad 
emission lines (BEL) are clearly detectable.  They provide information 
about the redshift as well as the energetics of the central engine and 
about radiative environment of the sub-parsec/parsec scale jets. 
It turns out that the energy density of the BEL region on sub-parsec/parsec 
scales as measured in the comoving frame of the jet usually dominates 
the energy density of the synchrotron radiation.  Within the framework 
of the internal shock model for the generation of radiative outbursts, 
that implies that the 1 -- 10 day time scale $\gamma$--ray flares 
are produced by Comptonization
of emission lines (external-radiation Compton [ERC] model) (Sikora, Begelman,
\& Rees 1994), rather than by Comptonization of synchrotron radiation 
(synchrotron-self-Compton [SSC] model), the latter presumably dominating 
the $\gamma$--ray production in the low luminosity BL Lac objects. 

The ERC model successfully predicts the observed location of 
the spectral break in the high energy spectra of FSRQs to be in the 1 -- 
30 MeV 
range without the necessity  of postulating a break in the power-law 
electron injection function.  According to the model, this spectral 
break simply reflects 
the break in the electron energy distribution caused by the effect of particle 
cooling (B{\l}a\.zejowski et al. 2000; Sikora et al. 2001; see also \S 2). 
If this interpretation is correct, the change of the spectral slope around 
the break resulting from cooling should be unique and  equal to
$\Delta \alpha_{x\gamma} = 0.5$, where 
$\alpha: F_{\nu} \propto \nu^{-\alpha}$.
In most FSRQs, $\Delta \alpha_{x\gamma} \le 0.5$, and the values of 
$\Delta \alpha_{x\gamma}$ lower than 0.5 
can be explained by invoking the possible dominant contribution of the 
SSC process in the soft/mid X-ray band.  However, the so-called  MeV-blazars 
present a challenge for the model.  Their X-ray spectra are very hard, 
$\alpha_x = 0.3 - 0.5$, and the high-energy $\gamma$--ray 
spectra are very soft, 
$\alpha_{\gamma} > 1.4$, and, therefore, $\Delta \alpha_{x\gamma} \ge 1$
(Blom et al. 1998; Bloemen et al. 1995; Tavecchio et al. 2000).
There were several attempts to explain such spectra in terms of
the ERC model.  Sikora et al. (1997) proposed an inhomogeneous model,
according to which the X-ray spectra are due to superposition of multiple 
ERC components produced at different distances in a jet and having low 
energy spectral cutoffs at frequencies determined by the plasma temperature, 
assumed to increase with a distance.  
This model, however, cannot reproduce the 
correlation of the  X-ray and $\gamma$--ray light curves during the 
1996 February flare observed in 3C~279 (Wehrle et al. 1998; Lawson et 
al. 1999).  Georganopoulos, Kirk, \& Mastichiadis (2001) suggested that 
the steep $\gamma$--ray spectra can result from the fact that they 
are softened by the Klein-Nishina effect.  But this effect is 
important in the EGRET band (30 MeV -- 3 GeV) only if energy of the 
seed photons involved in the ERC process is more than 10 times greater 
than the typical energy of the BEL photons.  Finally, B{\l}a\.zejowski 
et al. (2000) demonstrated that soft $\gamma$--ray spectra can result 
from superposition of two components, one produced by Comptonization of BEL, 
and another due to Comptonization of near-IR radiation produced by hot dust. 
This model requires the high energy cutoff of the ERC(IR) component
to be in the range 3 - 30 MeV and such a condition can be satisfied only via 
fine-tuning of the maximum electron energy to be within the range 
$3 \times 10^3 - 10^4$ $ m_ec^2$.
Another weakness of this model is that the spectra superposed from the
two components do not provide a good fit of the approximately power-law 
spectra observed in the EGRET band.

In the present work we assume that electrons are accelerated via a two-step
process and that their injection function takes the form of a double 
power-law with a break at the energy which divides the regimes of dominance 
of two different acceleration mechanisms.  Here the division energy is the 
threshold energy for the resonant scattering of electrons by Alfven waves 
(required to have efficient diffusive shock acceleration of electrons).  With 
this, we demonstrate that the MeV-blazar type spectra are produced when 
$\gamma$--rays result from Comptonization of the infrared dust radiation,
whereas the GeV-blazar type spectra  are produced if the $\gamma$--ray flux 
is dominated by Comptonization of BEL.  
Our work is organized as follows:  in \S2, we discuss the cooling effect
and demonstrate that for a distance range corresponding to the 
observed variability time scales, the cooling break is located 
within the 1 - 30 MeV range, as implied by observations.  In \S3, we 
discuss the issue of the electron acceleration efficiency and present the 
motivation for introducing the double power-law approximation for the 
electron injection function.  Such an approximation is used in \S4 to 
calculate the time-averaged spectra produced at different distances in 
a jet by shocks.  There, we also demonstrate that adopting the same electron 
injection function, one can reproduce the typical spectra of both MeV- and 
GeV-blazars. Our results are summarized in \S5.

\section{ELECTROMAGNETIC SPECTRA}
\subsection{Cooling effect}

The basic feature of the high energy spectra in FSRQs --- 
a spectral break between the X--ray and the $\gamma$--ray bands ---
has a natural explanation in terms of the ERC model.  In this model,
X--ray spectra are produced by electrons with radiative cooling time 
scale $t_{cool}'$, longer than the source life-time  $t_{fl}'$
({\it slow cooling regime}), whereas $\gamma$--rays are produced by 
electrons with $t_{cool}' < t_{fl}'$ ({\it fast cooling regime}).  
Noting that the angle-averaged cooling rate of electrons, when dominated by 
Comptonization of external radiation, is 
\be \vert \dot \gamma \vert \simeq {  \sigma_T  \over m_e c} u_{ext}' \gamma^2 \, , \ee
where $\gamma \equiv E_{el}'/m_ec^2$ is the random Lorentz factor of 
the electron and $u_{ext}'$ is the energy density of the diffuse external 
radiation field, both as  measured in the source comoving frame, 
we obtain the angle-averaged cooling time scale of the electron to be 
\be t_{cool}' \simeq {\gamma \over \vert \dot \gamma \vert} \simeq
{m_e c \over   \sigma_T}{1 \over  \gamma u_{ext}'}  \, , 
\label{eq:tc} \ee
where $u_{ext}'$ is the energy density of an external radiation field.
Then, from $t_{cool}' = t_{fl}'= {\cal D} t_{fl}$, where 
$t_{fl}$ is the observed time scale of the flare, and 
\be {\cal D}= {1 \over \Gamma(1 -\beta \cos{\theta_{obs}})} \, \ee
is the bulk Lorentz factor,
the break  in the electron distribution is located at the energy
\be \gamma_c \simeq {m_e c \over  \sigma_T} 
{1 \over u_{ext}' t_{fl} {\cal D}} \, . \label{eq:gac} \ee
For $\gamma < \gamma_c$, the slope of the electron distribution is
the same as the slope of the injection function; for 
$\gamma > \gamma_c$, the slope of the electron energy distribution is 
steeper by $\Delta s=1$
($s: N_{\gamma} \propto \gamma^{-s}$).

In the internal shock model, the lifetime of the flare 
is equal to the lifetime of the shock(s) and that is equal to the 
collision time scale of the two inhomogeneities, which prior to the 
collision are assumed to propagate down the jet with different velocities.
Thus, we have the distance range where the shock is active to be 
\be \Delta r_{coll} = c t_{fl} {\cal D} \Gamma \, ,\ee
and because time scales of flares in FSRQs are rarely shorter than 1 day, 
those flares 
are most likely produced at distances larger than 0.1 parsec.  At such 
distances, the contribution to $u_{ext}'$ is dominated by BEL 
and infrared radiation from hot dust. Noting that 
\be u_{ext}' = {1 \over c} \int I_{ext}' d\Omega' =
{1 \over c} \int I_{ext} {\cal D}_{in}^{-2} d\Omega \simeq 
u_{diff} \Gamma^2  \,  \ee
where 
\be {\cal D}_{in} = {1 \over \Gamma (1-\beta \cos \theta_{in})} \, \ee
and $\theta_{in}$ is the angle between the photon direction and the jet axis,
we predict  that  the break in an electron energy distribution at $\gamma_c$ 
should correspond to a break in the electromagnetic spectrum at frequency
\be \nu_{c} \simeq {\cal D}^2 \gamma_c^2 \nu_{diff}
\simeq \left({m_e c \over \sigma_T}\right)^2 
{\nu_{diff} \over u_{ext}^{'2} t_{fl}^2 }
\simeq \left({m_e c^2 \over \sigma_T}\right)^2 
{\nu_{diff} \over u_{diff}^2 \Delta r_{coll}^2} \, 
\left( {{\cal D} \over \Gamma} \right)^2 
\label{eq:nc} \ee 
and that the spectrum should change the slope around $\nu_{c}$ by 
$\Delta \alpha_{x\gamma} \simeq 0.5$ ($\gets \Delta s=1$).  

\subsection{External radiation fields}

\noindent
{\it Broad emission line region}

\noindent
According to the interpretation of the data obtained in many 
reverberation campaigns, production of broad emission lines in quasars is 
stratified and peaked around  a distance (see, e.g., Peterson 1993; 
Kaspi 2000; Sulentic, Marziani \& Dultzin-Hacyan 2000)
\be r_{BEL} \sim 3.0 \times 10^{17} \sqrt{L_{UV,46}} {\rm cm} \label{eq:rbel} \ee
Unfortunately, the detailed dependence of line luminosities on distance and  
geometry  of the BEL region is poorly known. 
For our illustrative  purposes we assume that the BEL region is spherical 
and that the fraction of the central UV radiation reprocessed into lines at
a distance $r$ is
$\xi_{BEL} (r > r_{BEL}) \equiv
\partial \xi_{BEL} /\partial \ln r \propto (r_{BEL}/r)^q$
and 
$\xi_{BEL}(r <r_{BEL}) = 0$.  With that, the energy density 
of BEL field, as measured in the comoving frame of 
the radiating plasma, can be approximated by 
\be u_{BEL}'(r) \sim  
{ L_{BEL} \Gamma^2 \over 4 \pi c r_{BEL}^2} \, 
{q \over 1 + (r/r_{BEL})^{2+q} } \, , \ee
where $L_{BEL} = \xi_{BEL} L_{UV}$ and
$\xi_{BEL} = 
\int_{r_{BEL}}^{\infty}(\partial \xi_{BEL} /\partial  r) dr$.
Note that for $r>r_{BEL}$, the value of $u_{BEL}(r)$ is dominated by 
radiation coming from smaller distances, from around  $r_{BEL}$.  However, 
since this radiation is redshifted in the source frame, the contribution to 
$u_{BEL}'(r)$ is dominated by the fraction of $u_{BEL}(r)$ which is 
determined by broad emission lines produced at a distance $r$.

\noindent
{\it Hot dust}

Evidence that hot dust is present in blazars is indirect.
Infrared emission of hot dust is directly measured in quasars which
are observed at larger angles to the jet axis than blazars. 
The ratio of the IR flux to the UV flux in these objects shows that the 
fraction of the UV radiation reprocessed by the dust into the infrared, 
i.e. the dust covering factor $\xi_{IR}$, is of the same 
order as $\xi_{BEL}$ (Sanders et al. 1989).  Dust is probably concentrated 
in molecular tori,
but optical extinction suggests that its distribution in the normal
direction to the equatorial plane  doesn't have any sharp boundary 
(Baker 1997).  One of the unknown aspects of dust in 
quasars is the minimal distance from the nucleus where it can exist.  
This distance can be limited by the maximum temperature that dust can survive, 
\be  r_{d,min} \sim {1 \over T_{d,max}^2} 
\left({  L_{UV} \over 4 \pi \sigma_{SB}} \right)^{1/2}  \ee
where $T_{d,max} \sim 1500$K,
or by the inner edge of the torus if it is larger than $r_d(T_{d,max})$ 
(see, e.g., Yi, Field, \& Blackman 1994). As observations of individual 
objects show, 
there is a very large scatter in the dust amount and its distance distribution 
amongst various quasars (Polletta et al. 2000; Andreani et al. 2002). 

In the present calculations we approximate the dust distribution by 
assuming that it is spherical and enclosed within
a given distance range with a constant covering factor.
With these assumptions the energy density  due to dust, as measured 
in the source comoving frame, can be estimated using the formula
\be u_{IR}'(r) \sim
{L_{d,IR} \Gamma^2 \over 4 \pi c r_{d,min}^2} \,
{1 \over 1 + (r/r_{d,min})^2} {1 \over \Lambda} \, , \ee
where $\Lambda =  \ln (r_{d,max}/r_{d,min}) \sim 5$.  The dependence of 
$u_{UV}'$ and $u_{IR}'$ on $r$ is illustrated in Fig. 1.  For such 
radiation fields we plot the dependence of $\gamma_c$ on $r$ on Fig. 2 
and of $\nu_c$ on $r$ Fig. 3.  It is apparent from the Fig. 3 that 
for a very wide range of $r$, the break $h\nu_c$ is located within 
the photon energy  range 1 MeV -- 30 MeV, in agreement with observations.

\subsection{Electron injection function}
As it was demonstrated by Sikora \& Madejski (2000), the energy flux
in powerful jets in quasars cannot be dominated by pair plasma.
This is because such a jet would produce much larger flux of 
soft X--ray radiation than is observed.  Hence, we assume that the 
inertia of the jet is dominated by protons.  In this case, the structure 
of shocks and the structure of Alfven waves generated around the shocks
are both determined by protons.  Being resonantly scattered by these waves,
the protons jump back and forth across the shock front and participate
in the 1st order Fermi acceleration process (Bell 1978;  Blandford \&
Ostriker 1978).  Those protons which  reach energies $> 10^9$ GeV interact 
efficiently with ambient photons and trigger (mainly via photo-meson 
process) synchrotron-supported pair cascades (Mannheim \& Biermann 1992). 
However, such a model fails to reproduce the very hard X--ray spectra of 
FSRQs (Sikora \& Madejski 2000).  This may indicate that in the context of 
those models, too few protons reach sufficiently high energies to power 
pair cascades.  

An alternative possibility is that the high energy radiation in FSRQs is 
produced by directly accelerated electrons.  However, for efficient
acceleration of electrons by the Fermi process via resonant scattering 
off Alfven waves, the electrons 
must be first pre\-hea\-ted\-/pre\-acce\-lera\-ted up to energies $\gamma_F$, 
at which point the magnetic rigidity of electrons becomes comparable with
magnetic rigidity of thermal protons, i.e. when their momenta are equal:
\be m_e \sqrt{\gamma_F^2 -1} \simeq  m_p \sqrt {\gamma_{p,th}^2-1} \, ,
\label{eq:gf} \ee
where 
\be \gamma_{p,th} -1 = \eta_{p,th} \kappa  \label{eq:pth} \ee
is the average thermal proton energy in the shocked
plasma, $\kappa$ is the amount of energy dissipated per proton in units of
$m_pc^2$, and $\eta_{p,th}$ is the fraction of the dissipated energy 
tapped to heat the protons. In the case of the two intrinsically identical 
inhomogeneities (Sikora \& Madejski 2001)
\be \kappa \simeq
{ ((\Gamma_2/\Gamma_1)^{1/2} - 1)^2 \over 2(\Gamma_2/\Gamma_1)^{1/2}} \, . 
\label{eq:kappa} \ee
where $\Gamma_2 > \Gamma_1 \gg 1$ are bulk Lorentz factors of inhomogeneities 
prior to the collision. 
For the reasonable assumption that 
$\Gamma_2/\Gamma_1 < 10$,  the thermal
proton plasma is at most mildly relativistic, i.e. $\gamma_{p,th}-1 < 1$.

Noting that 
\be n_e m_e (\bar\gamma_{inj}-1) \simeq n_p m_p  \eta_e \kappa =
n_p m_p{\eta_e \over \eta_{p,th}} (\gamma_{p,th} -1) \, , \ee
where 
$\bar\gamma_{inj}$ is the average Lorentz factor of injected electrons
and $\eta_e$ is the fraction of the dissipated  energy used to
accelerate electrons, 
we find that for  $\gamma_F$ and $\bar\gamma_{inj} \gg 1$
\be {\gamma_F \over \bar\gamma_{inj}} \sim 
{n_e \over n_p}{\eta_{p,th} \over \eta_e} \, 
\left({\gamma_{p,th}+1 \over \gamma_{p,th}-1}\right)^{1/2} \, .\ee
Hence, for non- or mildly relativistic shocks and for 
$\eta_{p,th} \sim \eta_e$, the threshold energy for the diffusive shock
acceleration  of electrons significantly exceeds the average 
energy of electrons, even if $n_e=n_p$.  This implies that the 
often-considered bulk preheating process is not able to provide 
an adequate number of electrons with $\gamma \ge \gamma_F$.  
However, as it was recently demonstrated in numerical PIC (particle-in-cell) 
simulations, the preheating/preacceleration mechanism can have a stochastic 
character and a non-negligible fraction of electrons can reach such 
energies (Dieckmann et al. 2000; Shimada \& Hoshino 2000).  Furthermore, 
the fact that even in blazars with the hardest X-ray spectra, the X-ray 
spectral indices $\alpha_X$ are always greater than 0 indicates that the 
largest number of electrons is injected at low energies, and, 
therefore, that there is no evidence for a bulk preheating process 
forcing most of electrons into equipartition with protons. 
This strongly suggests 
that in a similarity to the diffusive shock acceleration operating 
at $\gamma>\gamma_F$, the preacceleration process also has stochastic 
character and injects electrons with a power-law energy distribution.
Motivated by this, we assume that electrons, being accelerated by 
the two-step process, are injected  with the double power-law energy 
distribution
\be Q = \cases {C_l \gamma^{-p_l} & if $\gamma < \gamma_F$ \cr
C_h \gamma^{-p_h} & if $\gamma > \gamma_F$ \cr} \, , \ee 
where $C_h= C_l \gamma_F^{p_h-p_l}$. The injection break at $\gamma_F$ given
by Eqs. (13) - (15) results in a break in the electromagnetic spectrum at
the frequency
\be \nu_F \simeq {\cal D}^2 \gamma_F^2 \nu_{diff} \, .  \ee

\section{MeV-BLAZARS vs. GeV-BLAZARS}

For $2 <  \Gamma_2/\Gamma_1 < 10$ and $1/3 \le \eta_{p,th} \le 1/2 $, 
the break in the electron injection function is 
$\gamma_F \sim 10^3$ (see Eqs.~\ref{eq:gf} - \ref{eq:kappa} and Fig. 2).
Another break appears in the electron energy distribution at $\gamma_c$ 
(Eq. 4).
If the cooling of electrons is dominated by Comptonization of BEL, then
$\gamma_c \ll \gamma_F$, and the frequency range $[\nu_c,\nu_F]$
overlaps significantly with the EGRET band.  Hence, the spectrum 
in the EGRET band should show a slope $\alpha \sim p_l/2$ and should be 
steeper there than in the X-ray band by $\sim 0.5$ (or less, if 
the contribution of the SSC radiation into X-ray band is taken into account).
For $p_l \le 2.0$ such spectra become representative for short term
flares in GeV-blazars (Pohl et al. 1997).

At larger distances, the production of  $\gamma$--rays is dominated by
Comptonization of infrared radiation of hot dust.  Luminosity of that
radiation is comparable or even larger than the luminosity of the broad 
emission line light, but because of greater distance, its energy
density is much smaller than energy density of the diffuse radiation in 
the BEL region. As a consequence, the cooling break energy, 
$\gamma_c$, is now much larger, approaching the value comparable 
to the value of 
$\gamma_F$ (see Fig. 2).  Hence, there is no longer an extended spectral 
plateau with $\alpha = p_l/2$, and both $\nu_c$ and $\nu_F$ conspire 
to produce the spectral break $\Delta \alpha_{x\gamma} = 0.5 + (p_h-p_l)/2$.
Since now the energy of the Comptonized diffuse photons is 
$\nu_{IR}/\nu_{BEL}$ times lower than in the BEL region, an 
approximate location of this break is at a frequency which is by that factor 
lower than $\nu_F$ in the BEL region.  Hence, the spectra produced 
in the EGRET band at larger distances are predicted to have  
a spectral slope $\simeq \alpha_3 = p_h/2$.  For $p_l \le 2.0$
and $p_h \ge 2.4$ such  spectra become representative for MeV-blazars
(Tavecchio et al. 2000).

The above scheme can also explain two other observed differences between
MeV-blazars and GeV-blazars.  One is that the X--ray spectra are generally 
much 
harder in MeV-blazars.  In the context of the scenario above, 
this presumably results from the lower contribution of the SSC component to 
the X--ray band in the MeV-blazars than in the GeV-blazars, which is the case, 
e.g., if $(u_{IR}'/u_B')_{r_2} > (u_{BEL}'/u_B')_{r_2}$, where
$u_B'$ is the magnetic energy density and $r_2 \gg r_1 \sim r_{BEL}$.
The other difference is that in the spectra of MeV-blazars, in contrast 
to GeV-blazars, the thermal UV bumps are quite prominent 
(Tavecchio et al. 2000).  
This difference can be explained by noting that magnetic fields are weaker at 
larger distances, and therefore in MeV-blazars, the synchrotron spectra are 
shifted to lower frequencies, revealing the UV bump. All of the above 
effects are 
illustrated in Figs. 4 \& 5, where we present the time averaged spectra 
of radiation produced at two different distances by electrons injected
with the same energy distribution, $Q$, given by Eq. (18).
The spectra are computed using the internal shock model where the shock 
propagates down the conical jet, and including the following dominant 
radiation processes: synchrotron radiation -- SYN;
Comptonization of synchrotron radiation -- SSC; Comptonization of
BELs -- ERC(BEL);  and Comptonization of infrared radiation of hot dust --
ERC(IR).  The model is presented in B{\l}a\.zejowski et al. (2000)
and Sikora et al. (2001);  for a more comprehensive description of the model
see Moderski et al., in preparation.  The model input  parameters used to 
calculate the presented spectra are specified in the figure captions. 

In order to illustrate the relative location of the 
breaks at $\gamma_c$ and $\gamma_F$ in these two specific models,
we present in Figs. 6 and 7 the energy distributions of electrons.
They are computed at a distance of the shock termination (which in our models
is at $r=2r_0$) and presented together with the electron injection
function to demonstrate the cooling effect.  

It should be emphasized here that the value of $\gamma_F$ does not 
depend on how the fraction $(1-\eta_{p,th})$ of the 
energy dissipated in the shock is shared by other consistuents than 
thermal protons (see Eqs. 13 -- 15).  We also note that the comparison of 
the FSRQ spectra during outbursts against the spectra calculated from 
our model indicates that the best matching is provided by the models where the 
electron energy density exceeds the magnetic energy density by a 
factor ranging from a few up to more than 10. 

\section{CONCLUSIONS}

We demonstrated that the appearance of a FSRQ as a GeV-blazar or 
a MeV-blazar can be explained as depending on the distance of the maximum 
rate of energy dissipation in a jet.  The GeV-flat spectra are
predicted by our scenario to originate at distances where the production 
of $\gamma$--rays is dominated by Comptonization of BEL, 
while the spectra characteristic of MeV-blazars are predicted to 
originate at  distances where the production of $\gamma$--rays 
is dominated by Comptonization of hot dust radiation.  
An additional ingredient,
required to reach satisfactory unification between these objects, is
the assumption that electrons are injected with the double power-law energy
distribution, with the break at $\gamma \sim 10^3$.  Such location
of the break coincides numerically with the threshold energy
for the diffusive shock acceleration of electrons. Below that energy, the 
electrons must be accelerated by a different mechanism, e.g.
following instabilities driven by shock-reflected ions (Hoshino et al. 1992;
McClements et al. 1997; Shimada \& Hoshino 2000; Dieckmann et al. 2000);
or following reconnection of magnetic fields (Romanova \& Lovelace 1992; 
Blackman 1996).

As observations of X-ray spectra in MeV-blazars show, less energetic 
relativistic electrons are accelerated with the power-law energy 
distribution, with index $p_l \le 2$.  Future studies of dependence 
of the spectral shape on time scales of outbursts can be used to 
verify our idea that spectral breaks $\Delta \alpha_{x\gamma} >1$ result from 
superposition of two breaks, one produced by the cooling effect
and another one reflecting the break in the electron injection 
function at the threshold energy for the Fermi acceleration process.

It is very likely that some of the MeV-blazar phenomena can be
produced at distances where the source is already optically thin 
at mm-wavelengths.  If this is the case, the model 
predicts a correlation between variability
of the flux around the synchrotron-self-absorption break at mm-wavelengths 
and in the $\gamma$--ray band, on typical time scales of a month.
This possibility is supported by recent polarization  measurements 
at very high radio frequencies.  Those observations reveal 
that magnetic field in the very compact radio cores is dominated 
by its prependicular component and this is consistent with the model 
of the  radiation produced {\it in situ} by transverse shocks 
(Lister 2001). Depending on the amount and covering factor of the 
hot dust in the surrounding medium, 
such shocks can be sites of the MeV-peaked $\gamma$--ray blazars.  

Finally, we would like to comment that the MeV-blazar and GeV-blazar phenomena 
can interchangeably appear within the same object, as likely is the case 
in PKS 0208-512 (Stacy et al. 1996; Blom et al. 1996).  With such objects, 
correlation between the spectral type and the variability time scales 
can be studied directly, without the bias introduced, for instance, 
by the dependence 
of the variability time scales on the black hole mass which can be very 
different from object to object.  We note that the future sensitive 
$\gamma$--ray 
telescopes such as GLAST will provide us with excellent spectrally resolved 
light curves and will significantly constrain the applicability of 
radiative and particle acceleration models for those enigmatic sources.  


\acknowledgments

This project was partially supported by Polish KBN grant 5 P03D 002 21
and a Chandra grant from NASA to Stanford University via the SAO award 
no. GO0-1038A.
M.S. thanks SLAC for its hospitality during the completion of part of this 
work.


\clearpage

\centerline {\bf FIGURE CAPTIONS}

\figcaption[fig1.ps]{The dependence of the energy densities of external
radiation fields,
$u_{BEL}'$ (dashed line) and $u_{IR}'$ (solid line), and of energy density 
of magnetic field, $u_B' = B^{'2}/8\pi$ (dotted line) on the distance from 
the central source, $r$.
The curves are calculated using Eqs. (10) and (12) for: 
$r_{BEL}=5.2 \times 10^{17}$ cm; $r_{d,min}=2.6 \times 10^{19}$ cm;
$ \Gamma=15$; $L_{BEL}=3.0 \times 10^{45}$ erg s$^{-1}$; 
$L_{d,IR}=9.0 \times 10^{45}$ erg s$^{-1}$; $q=1.0$; $\Lambda=3.2$; 
and $B'=(1.2 \times 10^{18}{\rm cm}/r)$ Gauss.}

\figcaption[fig2.ps]{The dependence of the electron energy distribution
breaks, $\gamma_c$ and $\gamma_F$, on distance.
The $\gamma_c$ curves are calculated using Eqs. (4) and (5) for 
$r=\Delta r_{coll}$: $\gamma_c(UV)$ (dashed line) --
assuming that external radiation 
field is totally dominated by $u_{BEL}'$; and $\gamma_c(IR)$ (solid line)
-- assuming that external radiation field is totally
dominated by $u_{IR}'$.  The value of $\gamma_F$ 
is obtained from Eqs. (13) -- (15) for $\Gamma_2 / \Gamma_1=3.0$ and 
$\eta_{p,th}=0.5$.  
All other model parameters are the same as in Fig. 1.}
 
\figcaption[fig3.ps]{The dependence of the ``cooling break'', $\nu_c$
on the distance from the central source $r$. 
The break is calculated  for the same model parameters,
as in previous figures and assuming $\theta_{obs}=1/\Gamma$ 
($\to {\cal D}=\Gamma$).}
 
\figcaption[fig4.ps]{The model spectrum of the time averaged flare produced
by the shock formed at a distance $r_0 = 6.0 \times 10^{17}$ cm and terminated 
at $r=2r_0$. The shock propagates down the conical jet with a half-opening
angle $\theta_j = 0.02$.  The electrons/positrons
are injected at the rate given by Eq. (18) for $C_l= 3 \times 
10^{49}$ s$^{-1}$, $p_l=1.8$, and $p_h=2.8$.
All other model parameters are 
the same as in previous figures. The two areas confined by vertical dashed 
lines represent the 2 -- 10 keV and 30 MeV -- 3 GeV bands, respectively. 
The sharp feature around
$\nu \simeq \Gamma^2 \nu_{BEL}$ is the artifact of two approximations 
of our model:  first, that we do not take into account
upscattering of external photons produced at the distances smaller
than the actual position of the shock in a jet -- which would
contribute to spectrum at $\nu < \Gamma^2 \nu_{BEL}$, and 
second, that we use the average BEL frequency.}

\figcaption[fig5.ps]{Same as Fig. 4, but for the shock formed at a distance 
$r_0 = 2.0 \times 10^{19}$ cm.} 

\figcaption[fig6.ps]{Energy distribution of electrons $\gamma^2 N_{\gamma}$ 
at $r=2 r_0$ for $r_0 = 6.0 \times 10^{17}$ cm, shown as a dotted 
line.  For illustration, we also show the electron injection function 
multiplied by the period of the shock operation 
$\gamma^2 Q t_{fl}'$ as a solid line. }

\figcaption[fig7.ps]{Same as Fig. 6, but for the shock formation distance 
$r_0=2.0 \times 10^{19}$ cm. }

\clearpage\centerline{\psfig{file=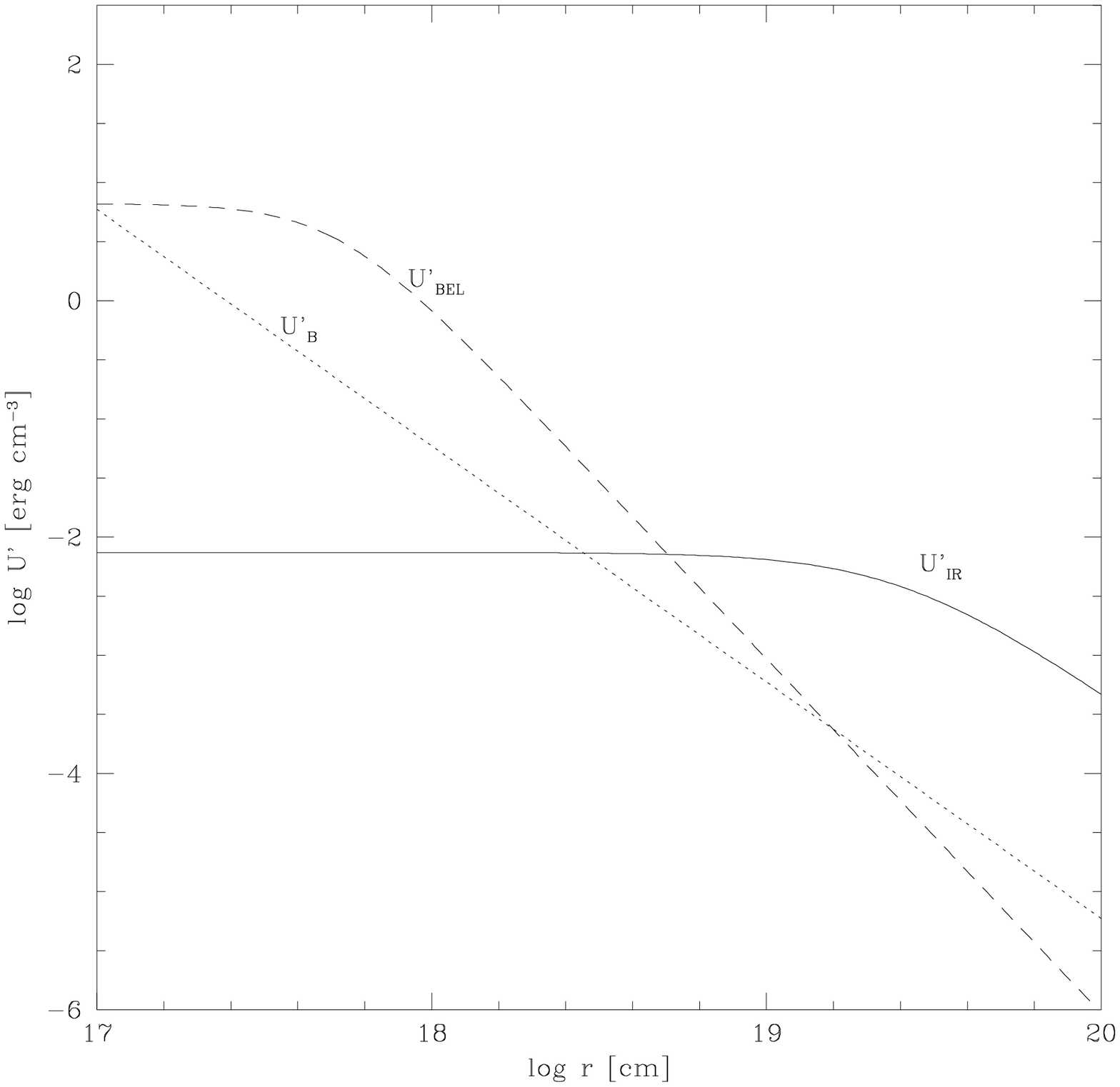,height=5.3 in,angle=0}}
\vfill\eject
\centerline{\psfig{file=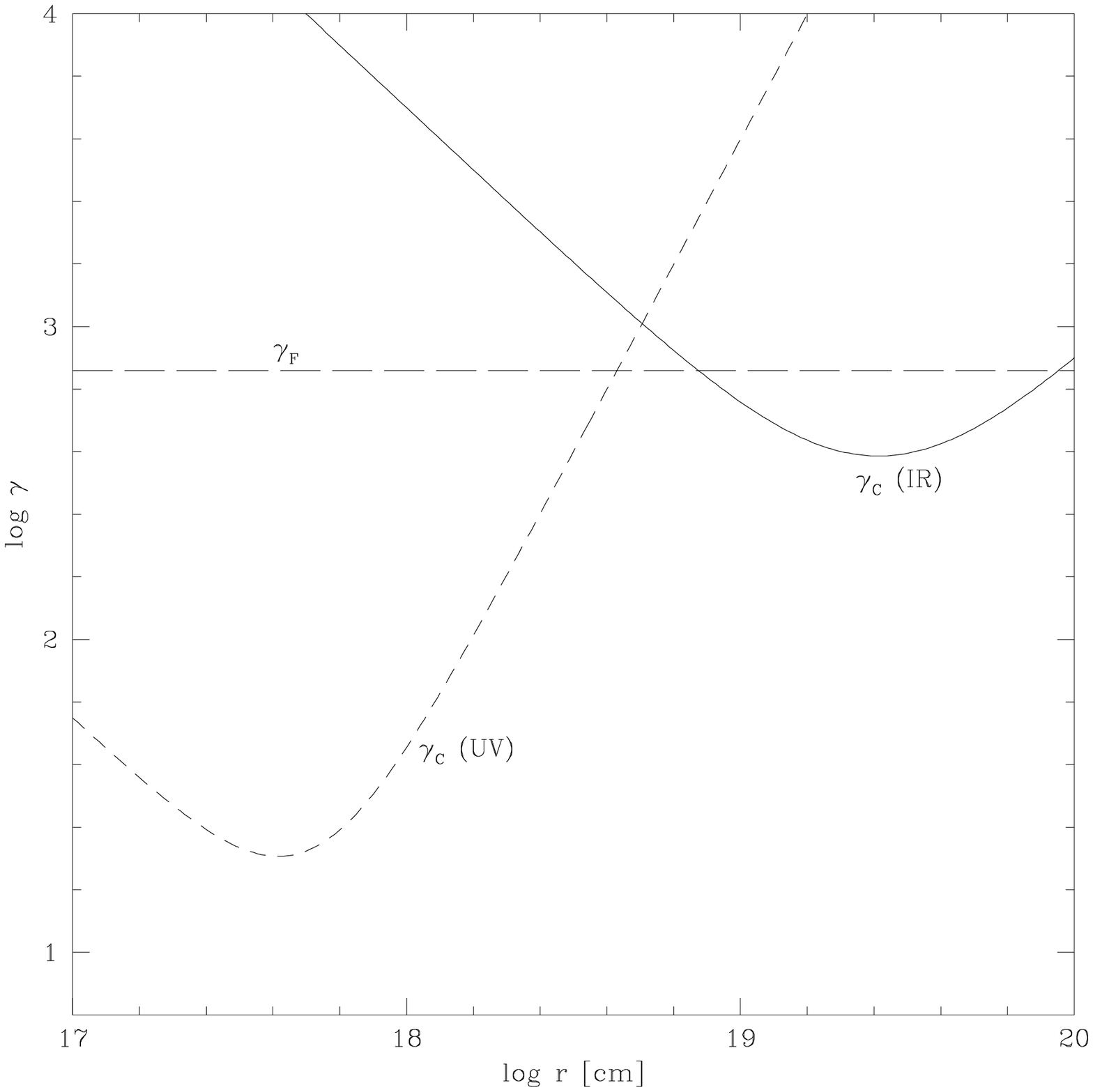,height=5.3 in,angle=0}}
\vfill\eject
\centerline{\psfig{file=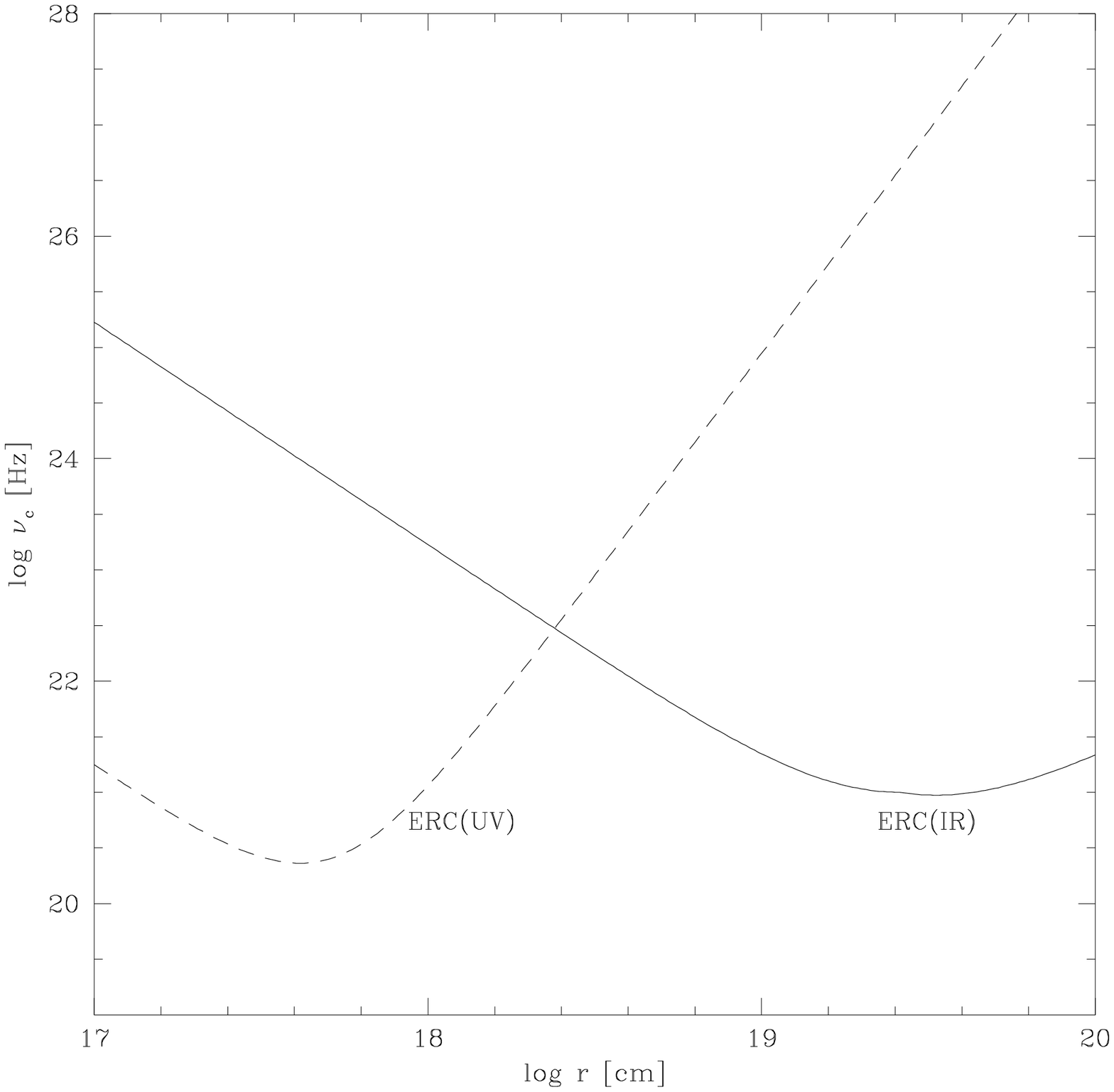,height=5.3 in,angle=0}}
\vfill\eject
\centerline{\psfig{file=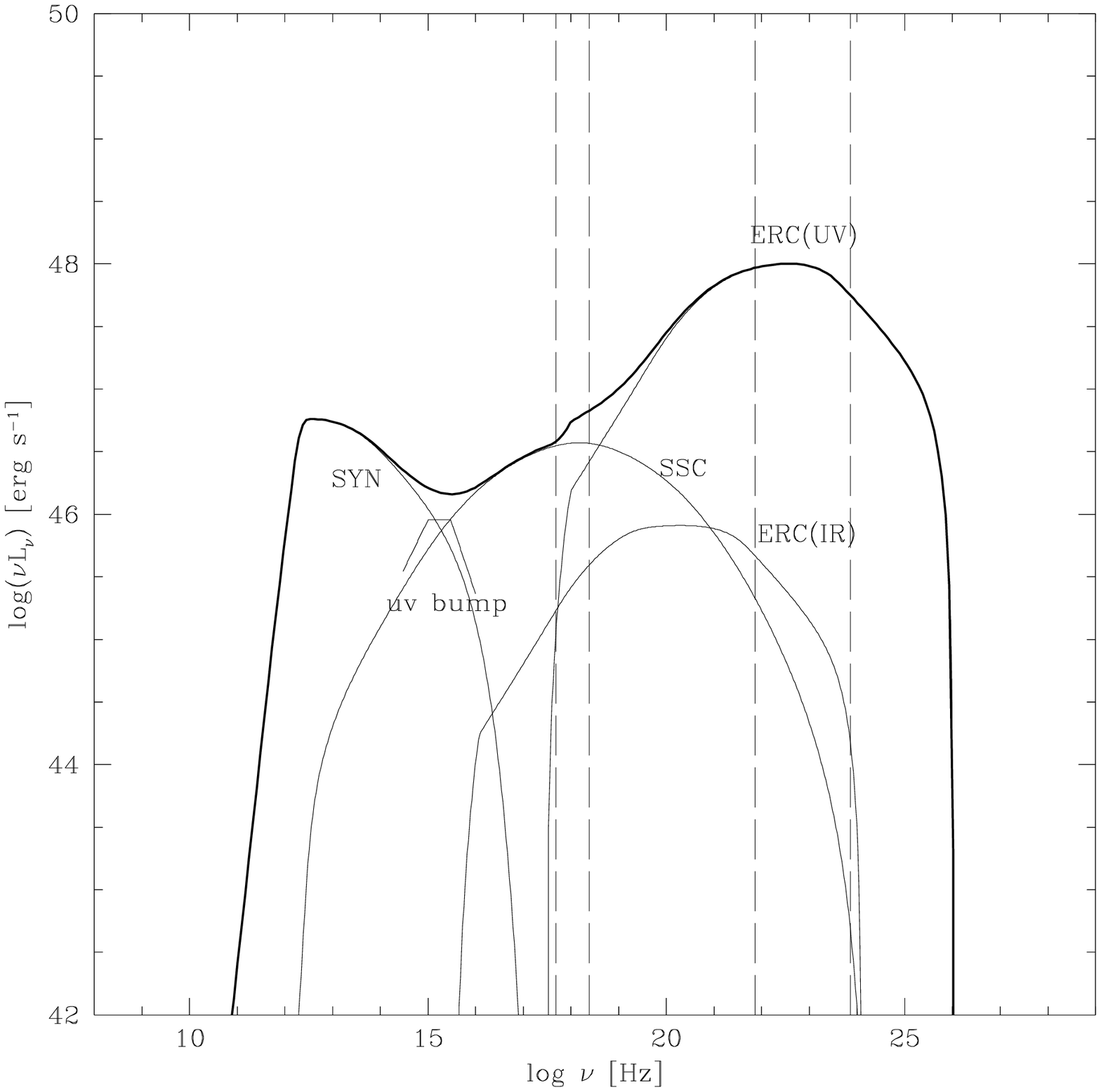,height=5.3 in,angle=0}}
\vfill\eject
\centerline{\psfig{file=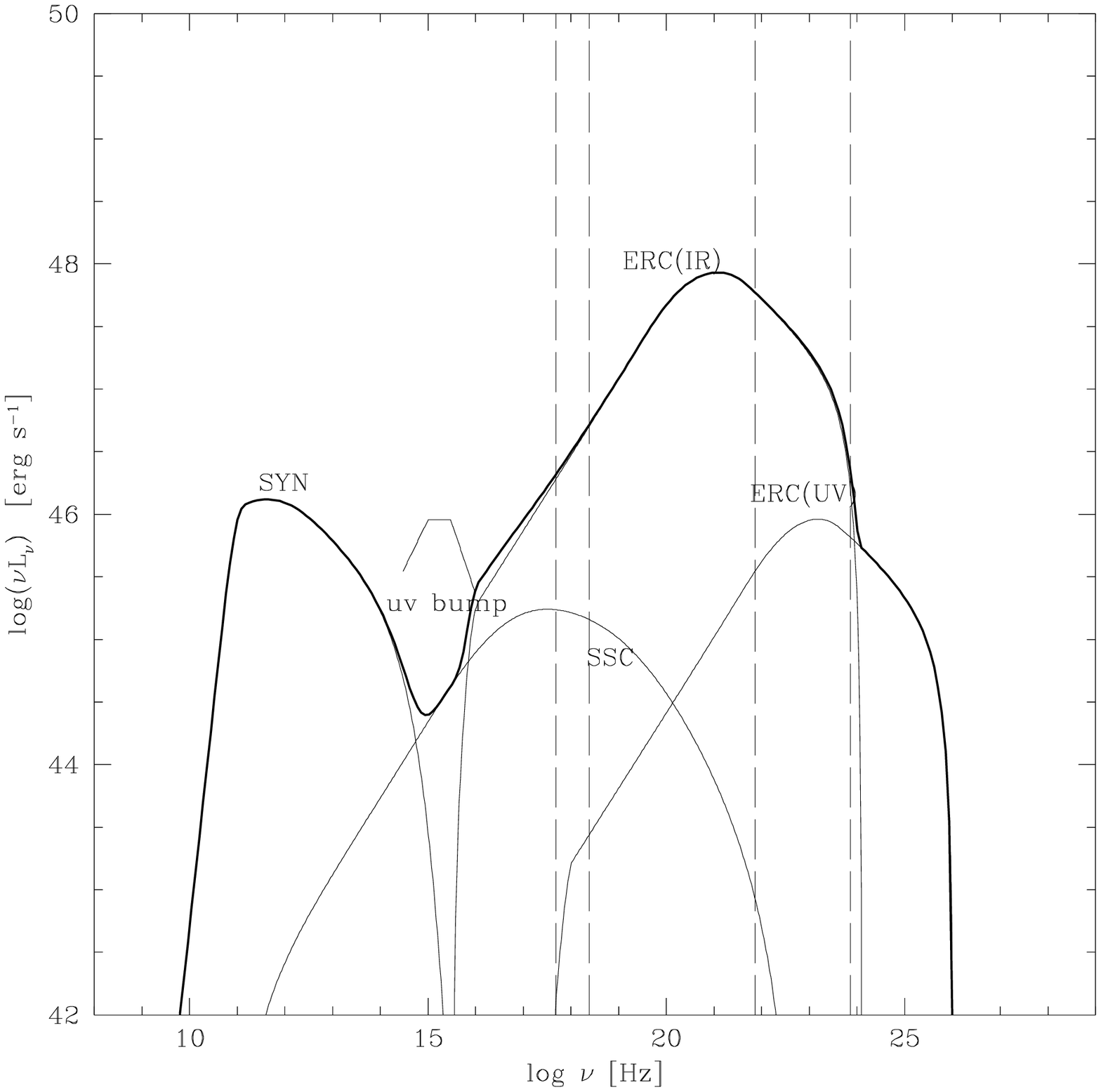,height=5.3 in,angle=0}}
\vfill\eject
\centerline{\psfig{file=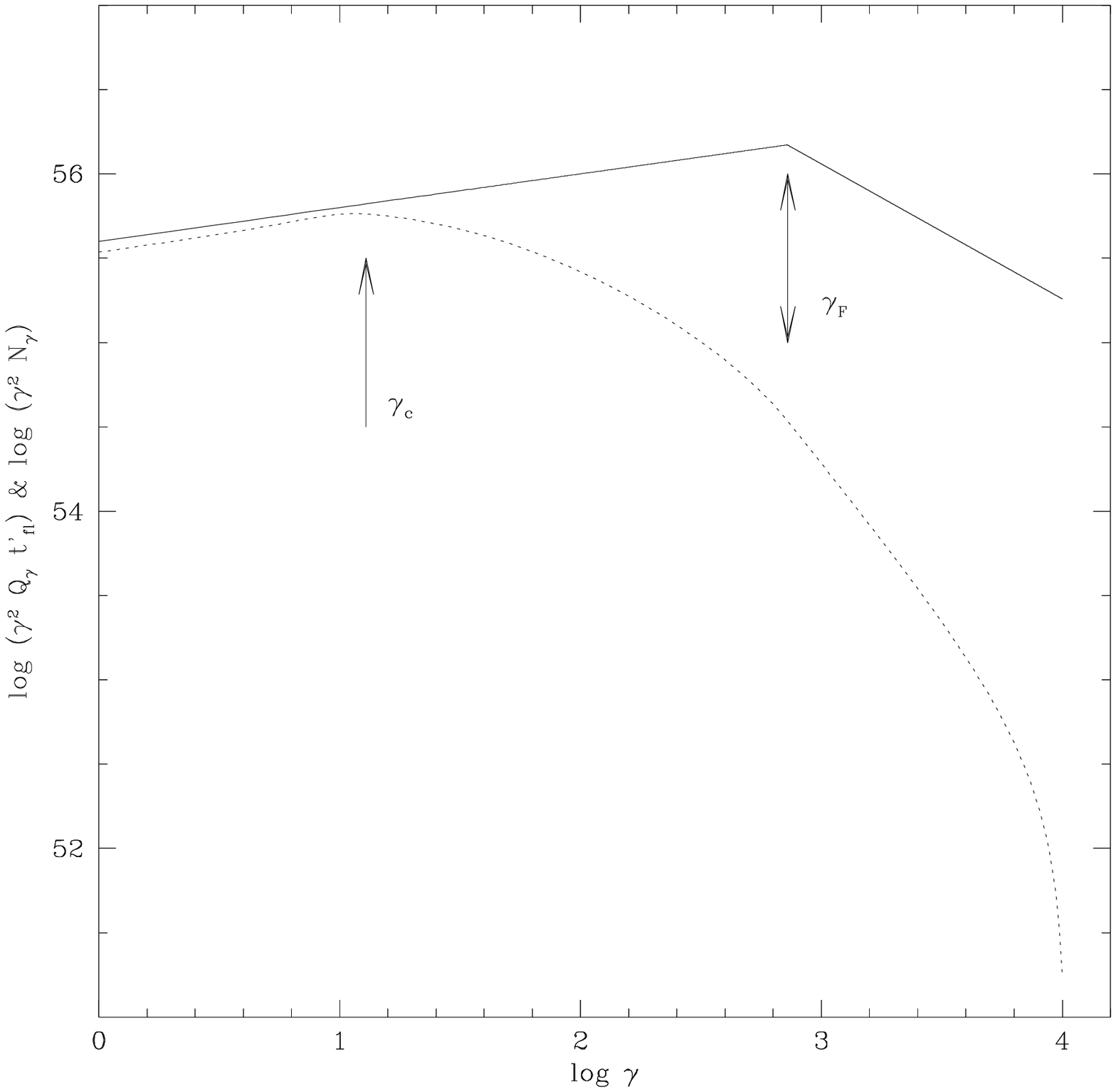,height=5.3 in,angle=0}}
\vfill\eject
\centerline{\psfig{file=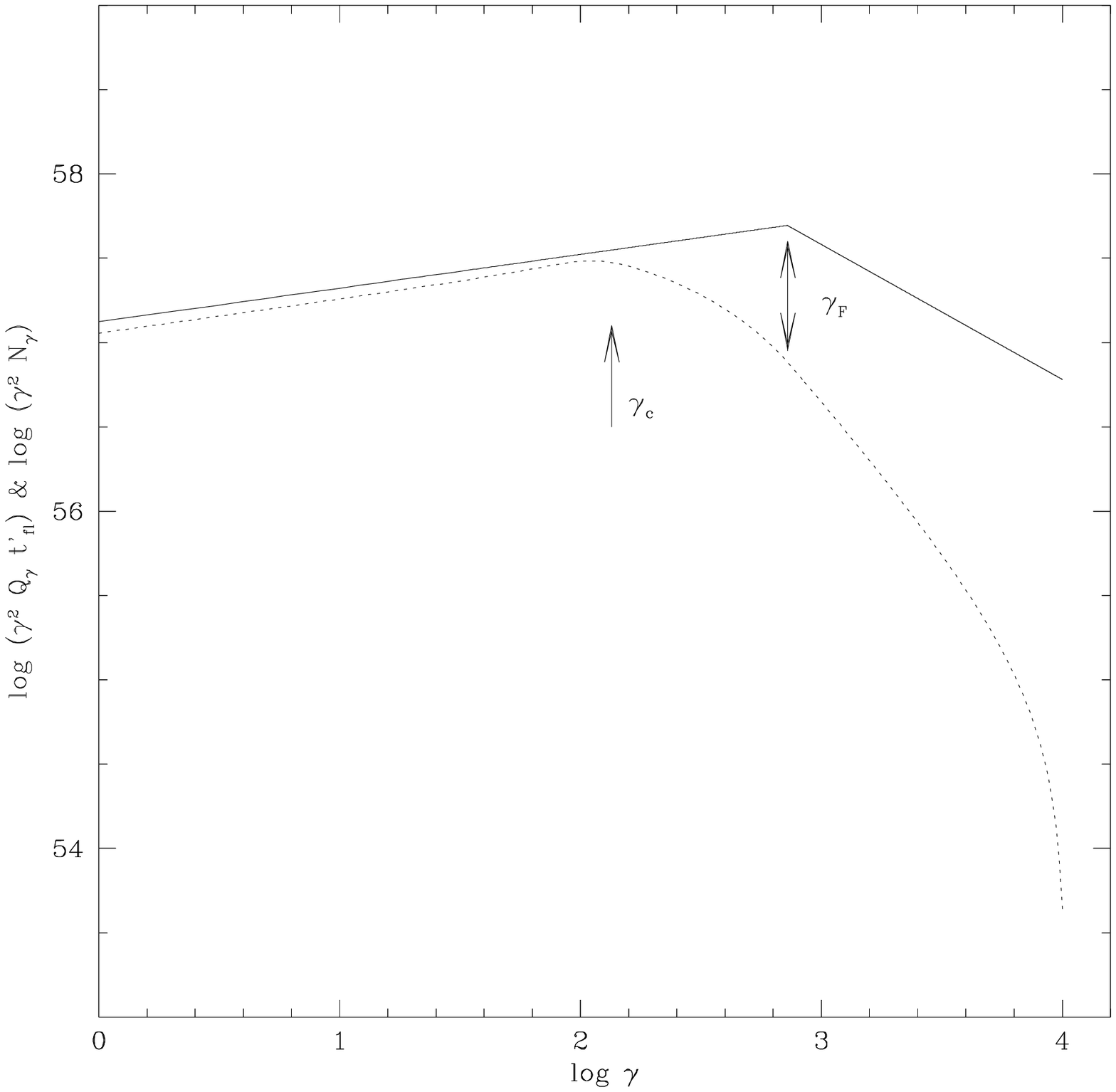,height=5.3 in,angle=0}}


\begin{thebibliography} {}
\bibitem{And} Andreani, P., Fosbury, R.A.E., van Bemmel, I., \& Freudling, W.
2002, A\&A, 381, 389
\bibitem{Bak} Baker, J.C. 1997, MNRAS, 286, 23
\bibitem{Bell} Bell, A.R. 1978, MNRAS, 187, 142
\bibitem{Blac} Blackman, E.G. 1996, ApJ 456, L87
\bibitem{BO} Blandford, R.D., \& Ostriker, J.P. 1978, \apj, 227, L49
\bibitem{BR} Blandford, R.D., \& Rees, M.J. 1978, in Pittsburgh Conf. on
BL Lac Objects, ed. A.M. Wolfe (Pitthsburgh: Univ. Pittsburgh Press), 328
\bibitem{Bla} B{\l}a\.zejowski, M., Sikora, M., Moderski, R., \&
Madejski, G.M. 2000, \apj, 545, 107
\bibitem{Bloe} Bloemen, H., et al. 1995, A\&A, 293, L1
\bibitem{Blom1} Blom, J.J., et al. 1995, A\&A, 295, 330
\bibitem{Blom2} Blom, J.J., et al. 1996, A\&AS, 120, 507
\bibitem{Die} Dieckmann, M.E., McClements, K.G., Chapman, S.C.,
Dendy, R.O., \& Drury, L.O'C. 2000, A\&A, 356, 377
\bibitem{Fos} Fossati, G., Maraschi, L., Celotti, A., Comastri, A., \&
Ghisellini, G. 1998, MNRAS, 299, 433
\bibitem{Geo} Georganopoulos, M., Kirk, J.G., \& Mastichiadis, 
A. 2001, \apj, 561, 111
\bibitem{Ho} Hoshino, A., Arons, J., Gallant, Y.A., \& Langdon, A.B. 1992,
\apj, 390, 454  
\bibitem{Kas} Kaspi, S. 2000, in ASP Conf. Proc. 224, Probing the Physics 
of Active Galactic Nuclei, eds. B.M. Peterson, R.W. Pogge, \& R.S. Polidan
(San Francisco: Astronomical Society of the Pacific), 347
\bibitem{Law} Lawson, A.J., McHardy, I.M., \& Marscher, A.P. 1999, MNRAS, 306,
247
\bibitem{Lister} Lister, M.L. 2001, ApJ, 562, 208
\bibitem{MB} Mannheim, K., \& Biermann, P.L. 1992, A\&A, 253,, L21
\bibitem{McC} McClements, K.G., Dendy, R.O., Bingham, R., Kirk, J.G., \& 
Drury, L.O'C. 1997, MNRAS, 291, 241
\bibitem{Muk} Mukherjee, R., et al. 1997, \apj, 490, 116
\bibitem{Pet} Peterson, B.M. 1993, PASP, 105, 247 
\bibitem{Pohl} Pohl M., et al. 1997, A\&A, 326, 51
\bibitem{Pol} Polletta, M., Courvoisier, T.J.-L., Hooper, E.J., \&
Wilkes, B.J. 2000, A\&A, 362, 75
\bibitem{RL} Romanova, M.M., \& Lovelace, R.V.E. 1992, A\&A, 262, 26
\bibitem{San} Sanders, D.B. et al. 1989, \apj,347, 29
\bibitem{SH} Shimada, N., \& Hoshino, M. 2000, ApJ, 543, L67
\bibitem{S94} Sikora, M., Begelman, M.C., \& Rees, M.J. 1994, ApJ, 421, 153
\bibitem[Sik01]{sik01} Sikora, M., B{\l}a\.zejowski, M., Begelman, M.C.,
\& Moderski, R. 2001, \apj, 554, 1 
\bibitem{SM} Sikora, M., \& Madejski, G.M. 2000, \apj, 534, 109
\bibitem{SM2} Sikora, M., \& Madejski, G.M. 2001, in AIP Conf. Proc. 558,
High Energy Gamma-Ray Astronomy, ed. F.A. Aharonian \& H.J. V\"olk
(Melville, New York: AIP), 275
\bibitem{SM3} Sikora, M., \& Madejski, G.M. 2001, preprint (astro-ph/0112231)
\bibitem{S97} Sikora, M., Madejski, G., Moderski, R., \& Poutanen, J. 1997, 
\apj, 484, 108
\bibitem{Sta} Stacy J.G., et al. 1996, A\&A Suppl.Ser.120, 549
\bibitem{Sul} Sulentic, J.W., Marziani, P., \& Dultzin-Hacyan, D. 2000,
ARA\&A, 38, 521
\bibitem{Tav} Tavecchio F., et al. 2000, ApJ, 544, L23
\bibitem{Mon} von Montigny, C., et al. 1995, \apj, 440, 525
\bibitem{Weh} Wehrle, A.E., et al. 1998, \apj, 497, 178
\bibitem{Yi} Yi, I., Field, G.B., \& Blackman, E.G. 1994, \apj, 432, L31
\end{thebibliography}
\end{document}